# The modification of the pore characteristics of activated carbon, for use in electrical double layer capacitors, through plasma processing

Marcelis L. Muriel[1], Rajaram Narayanan[2], Prabhakar R. Bandaru[1,2,3]

[1]Program in Materials Science, [2]Department of Nanoengineering,

[3]Department of Mechanical & Aerospace Engineering,

University of California, San Diego, La Jolla, CA 920393, U.S.A.

**ABSTRACT**

It was aimed to determine whether plasma processing could contribute to enhanced capacitance and energy density of activated carbon electrode based electrochemical capacitors, through the formation of additional surface charges. While an increase of up to 35% of the gravimetric capacitance, along with ~ 20% decrease in resistance, was obtained through optimal plasma processing, increased plasma exposure yielded a drastic reduction (/increase) in the capacitance (/resistance). It was also found that the capacitance and resistance modulation was a sensitive function of sample processing as well as electrochemical testing procedure. Considering the complexity of modeling realistic porous matrices, a metric to parameterize the reach of an electrolyte into the matrix has been posited.

**Text:**

The proliferation of portable devices and the related necessity for electrical energy provides a motivation for a deeper study of electrochemical energy storage[1,2]. The relative inefficiency of battery-related quick charging (*e.g.,* of use in portable electronics) or discharge (*e.g.,* for power spurts in automobile acceleration)[3] modalities has encouraged investigations into enhancing the energy density of electrochemical capacitors (ECs). In such devices, typically incorporating activated carbon (AC) electrodes[4–6], much of the energy (proportional to $CV^2$, with $C$ as the net obtainable capacitance over a voltage range: $V$) is stored on the surface for quick release, with the concomitant advantage of large power density (proportional to $V^2/R$ with $R$ as the series resistance of the EC). However, the attribute of charge storage only on the surface[7] also precludes large gravimetric and volumetric[8] energy density. In this paper, we consider a possible methodology for enhancing the energy density of AC electrodes. At the very outset, the material's relatively low cost, high surface area/mass ratio (>1000 m$^2$/g)[9,10], and ease of manufacture seem to be the underlying factors for its wide usage. However, the net obtainable capacitance obtainable from AC based electrodes is often hampered by inadequate electrolyte accessibility within the pores of the AC, in addition to space charge capacitance ($C_{sc}$), in series with the expected double-layer capacitance ($C_{dl}$)[11].

Previous work on nanoscale carbon (*e.g.,* carbon nanotubes[11–13] and graphene[14]) indicated that Argon plasma and related irradiation[15] could contribute to charge transfer and enhanced current density through atomic scale carbon removal and charged defect formation [12,16]. Ar plasma was chosen for its relatively simple constitution (*e.g.,* with Ar$^+$ as the only ion[17]), and relative lack of a tendency to introduce functional groups on the carbon surface[18], that may contribute to a redox capacitance[19,20] with reduced power density. Structural changes, as evidenced through Raman spectroscopy[12,14] as well as transmission electron microscopy indicated the creation of edge-plane defects, comprising electron rich dangling bonds, which have intrinsically higher density of states (*DOS*) and enhance the effective charge (and capacity).

In this paper, we aim to investigate whether such a capacitance increase would be facilitated through the plasma processing of AC. It is summarized that a 25% increase in capacitance could be obtained through the plasma processing.

**Experimental**

AC powder (Norit SX Ultra obtained from Sigma Aldrich Corp.) based electrodes were prepared without binder using 5% CNTs by weight, employing a vacuum filtration based process[21]. Briefly, 5 mg of CNTs were sonicated (using a Cole Parmer 8891 instrument) for 30 min in 100 mL of benzyl alcohol. Subsequently, 100 mg of AC powder was added to the solution, which was then sonicated for 1 hour. The solution was subject to vacuum filtration, and the filtrate washed with ethyl alcohol, and then deionized water. A porous carbon electrode, with nylon filter paper as the backing, was obtained with a typical mass loading of ~ 4 ± 1 mg/cm$^2$ after the filtration step. Surface area measurements via BET (Brunauer-Emmett-Teller) methods were performed on the untreated AC electrodes to confirm an internal specific surface area (SSA) of ~1000 m$^2$/g and a pore volume of ~ 234 cm$^3$/g. The typical thickness of the electrode was determined through Scanning Electron Microscopy (SEM) imaging: Figure 1(a), to be 200 ± 20 μm. The electrode was then dried at 100°C for 12 hours, and subsequently ~1 cm$^2$ pieces of the electrodes (comprising the AC as well as the filter paper) were used for electrochemical experimentation. The EC was then comprised of two AC electrodes (with filter paper backing) connected in series. In addition to the backing filter paper, ion-permeable membranes (from Dreamweaver Inc.) were also deployed. The fabricated electrodes were ion irradiated with argon plasma (in a Trion Reactive Ion Etching: RIE, chamber), to probe the effects of the plasma processing. The plasma power was set in the range of 100W - 500W, with an estimated ion density of ~ $10^{11}$ /cm$^3$[17,22].

The electrodes were tested via a two-electrode set-up (assembling apparatus from MTI Corp.): Figure 1(b). A robust and reliable low resistance contact, and abundant electrolyte (acetonitrile with dissolved Tetra-butyl-ammonium hexafluorophosphate: TBAHFP, at 1 M concentration), was used. The samples were subject to detailed electrochemical characterization, involving Constant Current Charge/Discharge (CCCD), Cyclic Voltammetry (CV), and Electrical Impedance Spectroscopy (EIS). The same sample (with a given AC mass loading) was used for testing under all the considered electrochemical tests.

An initial estimate of the AC electrode capacitance was obtained through standardized [23] CCCD based testing. The electrode was charged/discharged at a constant current ($I$) and the ratio of the applied current to the consequent rate of change of voltage would be a measure of the capacitance value (Eqn. 1)[23]. It was noted that at the beginning (/end) of a charge (/discharge) cycle, that there would be a voltage drop (= $IR_{CCCD}$) proportional to an equivalent series resistance (ESR) of the fabricated AC electrode based capacitor, and was correlated to the condition at the electrode-electrolyte interface. The double-layer capacitance ($C_{dl}$) was computed through

$$C_{dl} = \frac{I}{dV/dt} = \frac{I \cdot (\Delta t)}{\left(V'_{max} - \frac{1}{2}V'_{max}\right)} \qquad (1)$$

. The $dV/dt$ is the slope of the discharge curve, $V'_{max}$ is the maximum voltage after the $IR$ drop and $\Delta t$ corresponds to the time needed to achieve the voltage difference. The above definitions are in accordance with the prescriptions of a best-practice method[23].

The CV technique considers the variation of the electrical current in the AC electrodes as a function of applied voltage, over various scan rates, with the area enclosed by the resulting characteristics proportional to the power that can be obtained from the AC electrode in the EC, and which when multiplied by the time yields the energy density. In our experiments, the CV was conducted at various voltage scan rates ($v$) in the range of 10 mV/s to 500 mV/s. The double layer capacitance ($C_{dl}$) and the gravimetric capacitance ($C_{gm}$) was estimated from the net area of the resulting voltammogram [23] through:

$$C_{dl} = \frac{P(t)}{v \cdot V(t)} = \frac{\int_{V_1}^{V_2} i\, dV}{2 \cdot v \cdot V(t)} \ldots\ldots\ldots (2a) \qquad C_{gm} = \frac{2 C_{dl}}{A_{proj} M_{ave}} = \frac{\int_{V_1}^{V_2} i\, dV}{v \cdot V(t)} \ldots\ldots\ldots (2b)$$

Here, $P(t)$ is the power in the EC, with $V_1$ and $V_2$ as the initial and final voltage limits, respectively, of a voltage cycle conducted at a given $v$, yielding a time varying voltage: $V(t)$ and electrical current: $i$. The electrode geometry and mass is considered through the projected area ($A_{proj} \sim 1$ cm$^2$ in the reported experiments), with $M_{ave}$ as the average mass loading (~4 mg/cm$^2$). Since the two AC electrodes in the cell are in series: Figure 1(b), the capacitance would be twice that of the calculated value: Eqn. 1(b).

We also employed EIS[24,25], where the impedance of the AC electrode was determined as a function of the alternating current frequency ($f$) in the range of 10 mHz to 300 KHz, with an imposed constant DC bias (of 1 or 2 V). Both the bulk electrolyte based equivalent series resistance ($R_s$), the charge transfer resistance ($R_{ct}$), the diffusion based Warburg impedance: $Z_w$, in addition to the $C_{dl}$ was estimated. Additionally, the resistance due to the inter-granular contacts between the particles in the electrode [26] was also incorporated to carefully model EIS characteristics.

**Results and Discussion**

We report on the results of the electrochemical testing of the pristine/untreated and plasma processed AC electrodes. At the very outset, as indicated in Table I, it was noted that (i) there seems to be an optimal plasma processing treatment for a capacitance increase, with increased plasma power decreasing the capacitance, and (ii) there was considerable diversity in the obtained capacitance values, specific to the employed electrochemical technique. In this section, in addition to the characterization of such features, we aim to provide a rationale for comparison between the values obtained by the various techniques.

The CCCD measurements were used for initial estimates of the capacitance, from Eqn. (1). As seen from Figure 2, the shape of the voltage-time plots was generally linear. From the inverse proportionality of the capacitance to the slope of the plots, an optimal plasma processing was inferred, *e.g.,* the processing at 100 W for 2 minutes seems to yield a higher capacitance compared to that at 500 W for 2 minutes. A lower time of charge (/discharge) was apparent for the 500 W plasma power processed sample. The experimentally deduced resistance ($R_{exp} = R_{CCCD}$) and capacitance ($C_{exp} = C_{CCCD}$) data indicated a definitive modulating influence of the plasma processing on the pore size distribution of the AC electrode, the product of which could be related to the times of residence of the electrolyte in the pores.

It was also clear through CV, employing Eqn. (2) that there seemed to be optimal plasma processing that was effective for increasing the capacitance and energy density. For instance, a capacitance increase (/decrease) was noted for the 100 W (/500 W) plasma processing: Figures 3(a) and 3(b). The increase of the curve past ~ 2.7 V in Figure 3(a), is possibly due to redox reactions close to the stability limit of the electrolyte or due to parasitic redox couples, *e.g.,* due to absorbed ionic species. Moreover, the shape of the curve was sensitive to the $v$ as well as the plasma processing power, with a more rectangular shape (typical of double-layer capacitors) being obtained at lower *v*, and at 100 W, *cf.,* Figures 3(a) and 3(b) compared to Figure 3(c). The beneficial influence of optimal plasma processing was further noted in Figure 3(d), which indicates a general, and expected, decrease of the gravimetric capacitance with increasing scan rate. A higher *v* is rate limiting in terms of diminishing the access of the electrolyte ions into the pores and generally reduces the observed capacitance, as in Figure 3(d).

The rationale for the capacitance enhancement, at optimal plasma processing (*i.e.,* at 100 W for 2 minutes) seen in both CCCD and CV may be due to an effective increase in the area of charge storage, implying possibly a (a) modulation of the pore characteristics, or (b) due to additional charged defects, as was recently observed for graphene-like structures[14]. Observations from plasma processing (*e.g.,* done at 500 W, 2 minutes) seem to imply the former characteristic as more likely. However, it should be noted at the very outset, that the precise origin and nature of the enhanced capacitance is unclear. Generally, the extraction and modeling of a double-layer capacitance is non-trivial for porous electrodes assembled through powder consolidation, considering that the effective area of charge storage is not easy to determine due to the difficulty in assessing the actual accessibility by the electrolyte. For instance, while the current would be expected to increase with the *v*, from $I = Cv$, it was observed that the proportionality was not strictly obeyed. Indeed, the influence of ion penetration into the pores, as a function of the ion approach frequency as well as the pore size and shape is complicated[19,27]. Consequently, the overall impedance of a porous electrode system would be aggregated in from the influences of the expected characteristics for a flat electrode compounded with a *distributed*/equivalent series resistance (*R*) and the capacitance as related to the pore constitution of the AC electrode. The consideration of such a distributed resistance leads to, for example, the actual voltage ($V_{actual}$) varying with the $R$ and time ($t$), through: $V_{actual} = V_{app} \pm vt - IR$, where $V_{app}$ is the applied voltage and $I$ is the capacitor charging/response current. Consequently, the charging current was modeled to be of the form[28]:

$$I = vC - \left(\frac{2vC}{1+\exp\left(-\frac{V_{max}}{vRC}\right)}\right)\exp\left(-\frac{t}{RC}\right) \tag{3}$$

$V_{max}$ is the maximum voltage reached during the CV scan and the experimental resistance $R$ (= $R_{CV}$) was determined from a fitting of the obtained CV curves (*e.g.,* in Figure 3(a), (b), and (c)). The above form of the current yields a decreasing current at increasing $v$ as well as a transition to an elliptical form[28]: Figure 3(c). A good correlation of the determined $R_{CV}$ to that obtained from the ratio of the power (area of the voltammograms)/$(V_{max})^2$ was seen. A closer examination of the CV curves in Figures 3(b) and (c), also seems to indicate a higher resistance manifest at the electrode-electrolyte interface at larger $v$ for the samples subject to plasma processing at 500 W @ 2 minutes, and may be implicated in a reduced current at a given voltage. Alternately, the current is increased for samples plasma processed at 100 W @ 2 minutes, indicating lower resistance. Such resistance increase or decrease could be interpreted through pore widening or closing phenomena, and would be better understood through a more detailed interpretation of the EIS: Figure 4, which will be discussed next.

While the EIS technique, in principle[24], may be used to delineate many characteristic impedances (both resistances and capacitances), it is often difficult to attribute specific features of the resulting plots to physical characteristics of the porous electrodes. The presence of a semi-circular loop at high frequencies, as observed in the plots of Figure 4, has previously been ascribed to additional resistance arising from the ionic conductivity limitations, due to "intergranular contacts" [26] and has been referred to as a *pseudo-transfer* resistance[29], with a larger radius of the semicircle correlated to a larger micro-pore volume. At the highest frequencies, considering the intercept on the real axis of the Nyquist plot (Figure 4(a), also see the *inset*), it was noted that the series resistance (ESR: $R_s$) is generally increased due to the plasma processing. It was also observed that while the *untreated* and the sample processed at 500 W @ 2 minutes exhibit related impedance spectra at high frequencies (> 1 kHz), the sample with *moderate* plasma processing: at 100 W @ 2 minutes, exhibited a forward shift of the $R_s$ (at the highest frequency of 300 kHz) as well as a depressed semi-circle. The individual EIS plots for the various samples, *i.e.,* untreated: Figure 4b, 100 W @ 2 minutes: Figure 4c, and 500 W @ 2 minutes: Figure 4d, at two representative bias voltages (1V and 2V) are also shown.

It was previously noted that the distributed resistance and capacitance of a pore distribution generally leads to a smaller value of the magnitude of the measured impedance[30]. It was discussed [29] that Norit based carbons, as used in the present study, generally have a mesoporous distribution, with enhanced pore accessibility. However, there generally seems to be a substantial dispersion of the capacitance mimicking diffusional control behavior, in the low frequency regime in most cases, due to a pore size distribution. The specific shape of the pore was intimately connected to the impedance behavior in the medium frequency regime[19,27].

The EIS variation could be interpreted in terms of the effects of the plasma processing on the *average* pore geometry. As indicated previously[19,27], and suggested in Figure 4(a), a decreasing pore diameter at the entrance with respect to the overall pore size leads to a greater extent of depressed/reentrant behavior of the *Z"*. It may then be speculated that the *untreated* and the 500 W processed sample both have equivalent pore structure effects, and dissimilar to that observed for the 100 W plasma processed sample. However, the 500 W plasma processed sample has much greater capacitance dispersion in the low frequency regime, *i.e.,* due to large deviation from a perpendicular line and closely related diffusive limitations. A larger voltage (*i.e.,* 2 V compared to 1 V) also leads to such considerations, as in Figures 4(b), (c), and (d).

The conclusion thus far seems to be that plasma processing modifies the geometry of an effective pore. However, while this seems to yield some insight, there still seems to be some necessity for *ad hoc* assumptions to justify the obtained data. For instance, the EIS technique probes impedance over a range of frequency (/time) scales, from ~ 10 mHz (~100 s) to ~ 300 kHz (~3 µs), implying electrolyte interactions over such scales. However, even considering both *in-a-pore* and *by-pore* size distributions [31], where the electrolyte penetration decreases with increasing frequency - at the same pore radius, or decreases with decreasing radius – at the same frequency, respectively, does not yield much insight due to the complexity of realistic pore distribution. Many models, which are primarily analytical in origin and rather simplistic, *e.g.,* assuming either a pore of uniform geometry (*i.e.,* a cylinder of constant length or radius[30,32]) or pores following a deterministic probability distribution function[31], following transmission line precepts, have been used. Many such models also use a lumped element approach, delineating the resistances and capacitances per unit length, which are often supposed to be uniform along the length of the pore. The AC electrode performance, for example in EIS, is related in a complicated way to the *overall* capacitance and resistance and while conventionally modeled in terms of a $C_{dl}$ and the analogous resistances, *i.e.,* the series resistance ($R_s$) and the transfer resistance ($R_t$), many other parameters *specific* to a porous electrode (*i.e.,* $R_p$ and $C_p$) must also be considered. However, the $R_p$ and $C_p$ are invoked rather superficially and fit to a specific model to explain away non-conventional attributes of the capacitance of the porous electrodes. Other issues, such as those related to impedance matching and frequency dispersion (as would occur for a constant applied voltage/current) are well known[33]. However, as originally stated more than five decades ago, such detailed models "would not be of much use" [32] due to ambiguity related to determining an "effective pore diameter …or an effective electrode surface". While we essentially agree with the above statement, reductionist approaches may still yield some insights. We propose a metric that could indicate the relative efficiency of porous electrodes with respect to electrolyte accessibility and the consequent modulation of the observed resistance and capacitance.

We present an alternative treatment that goes beyond the single pore or pore distribution-based approaches, in terms of considering the *reach* ($\zeta$) of an electrolyte into a porous electrode. We do not assume *a priori* any specific pore geometry (length or diameter) or pore size distribution. At the very outset, the fundamental assumption is that the $\zeta$ would be a function of *both* the experimental time (say, $t$) as well as a time constant ($\tau$) equivalent to the product of the probed/determined capacitance ($C_{det}$) and resistance ($R_{det}$) of the porous electrode, *i.e.,* $\tau = R_{det}C_{det}$. The $\zeta$ is parameterized by the quantity $\sqrt{t \cdot \tau}$, estimated through a specific experimental determination (*e.g., det = CCCD, CV,* or *EIS*).

For instance, in CCCD, the relevant $\zeta_{CCCD} = \sqrt{t_{CCCD} \cdot \tau_{CCCD}}$ was obtained from the product of the $t_{CCCD}$, the time of discharge (from 2. 7 V to 0 V) and $\tau_{CCCD} = R_{CCCD}C_{CCCD}$, with the $R$ and the $C$ obtained through standard procedures, from Figure 2, considering the *IR* drop and the slope of the *V-t* characteristic, respectively. The correspondent values are all indicated in Table II. In the case of CV, at $v$ = 0.1 V/s, the experimental time ($t_{CV}$) to charge the activated carbon constituted capacitor from 0V to 2.7 V, *cf.* Figure 3(a), would be 27 s. The $R$ (= $R_{CV}$) and the $C$ (= $C_{CV}$) obtained through the fitting of the experimental curves to Eqn. (3), would yield the $\tau_{CV} = R_{CV}C_{CV}$. The $\zeta_{CV} = \sqrt{t_{CV} \cdot \tau_{CV}}$ and is depicted in Table II for the unprocessed and plasma processed carbon electrodes. While the EIS involves probing eight orders of magnitude in frequency/time scales, a comparison to the values estimated from CV may be tentatively rationalized through considering the time scales at the lowest frequency (of $f$ = 10 mHz corresponding to ~ 105 seconds, in our experimental setup). Consequently, we assumed a corresponding value for the $t_{EIS}$ while the $R_{EIS}$ was estimated from the linear extrapolation of the low frequency curve to $Z'' = 0$ and the $C_{EIS}$ was set to be = $1/j\omega Z''$, with $\omega = 2\pi f$. With $\tau_{EIS} = R_{EIS}C_{EIS}$, the computed values of $\zeta_{EIS} = \sqrt{t_{EIS} \cdot \tau_{EIS}}$ are shown in Table II.

It was observed, from Table II, that while the CV yields similar values of the $\zeta_{CV}$ for the untreated and plasma processed samples (at ~ 13.5 s), a range of $\zeta_{CCCD}$ and $\zeta_{EIS}$ times were obtained. It may also be inferred that the parameterization in terms of an effective $\zeta$ would be useful in terms of denoting the extent to which the porous matrix is *effectively* probed and may yield insights into the reach of the electrolyte in a given testing procedure. Additionally, the CV testing procedure seems to be adequate in that the $\zeta_{CV}$ is identical for all the samples while a range of $\zeta$ is obtained through the CCCD and the EIS tests, implying a variation of the pore characteristics as a function of plasma processing, as indicated in Figure 5. The basis pore shape related to the untreated electrode: Figure 5(a), was adapted through relating the corresponding EIS spectral features [27]. Given the relatively low powers of plasma processing and the low penetration of ions, *cf.,* ion implantation techniques, the surface pore morphology may be the most affected.

We have suggested that the testing procedures probe different characteristics of the porous matrix. We observe, *e.g.,* from Table II, that $C_{CV}$ decreases considerably, by a factor of three, with enhanced plasma processing (*e.g.,* 500 W @ 2 minutes) accompanied by a corresponding *increase* in the $R_{CV}$. While the decreased capacitance was also observed in the CCCD and EIS experiments, the resistance was found to be relatively constant. However, the determined resistance value for the AC electrode subject to 100 W of plasma power, generally shows a decreased resistance across all testing techniques, implying a larger pore entrance area: Figure 5(b). Such variations seem to indicate the sensitivity of the pore morphology, which may change as a function of plasma processing to the particular testing technique. For instance, in CV testing, with gradually increasing voltage over a fixed time, the ion penetration effectively samples the pore distribution and the increased resistance for the 500 W plasma processed sample may occur due to pore blockage effects, as indicated in Figure 5(c). The reduced capacitance would then be a consequence of a fewer number of ions reaching past the blockage region. A similar ion starvation effect is seen in EIS as well, albeit with a different cause. As EIS is carried out with a DC bias voltage of 1 V (on which is superimposed a much smaller AC voltage), the DC bias may increase the mean distance of the ions (the number of which have been considerably reduced due to the blockage), past the blockage region. The consequent reduction of the number of ions as well as the sampling of the pore beyond the blockage, yields a reduced capacitance and a relatively constant resistance. While the EIS then seems to probe further into the porous region, the CCCD testing samples pore surface effects. As CCCD is carried out under a fixed current, the relevant number of ions is constant. Consequently, the ions sample the pore surface proportional to their number. As indicated earlier, ion starvation effects beyond the pore blockage region reduce the capacitance to a very small value of ~ 22 mF – the smallest obtained in the three techniques. It is relevant to note that the resistance does not seem to be significantly affected implying that the blockage of the pore is not being considered in CCCD (as well as in EIS). Qualitative insights may be deduced through the $\zeta$ (or $\tau$) determination, *e.g.,* it may be expected that the *net* time scales for probing the porous matrix through CV would be lower than that through EIS, due to the alternating nature of ion flow into/out of the pores.

It is interesting to note that while Table I indicate a modulation of the gravimetric capacitance, an additional consideration of the resistance yields further insight into the structural changes related to the modulation. While the specific capacitance (*e.g.,* in F/g) still changes, as evident through data presented in Figures 2-4, the consideration of the $\zeta$ would yield insights into the relative influences of the testing time and the time required to effectively charge the capacitance of the porous electrode/s. A $\frac{\zeta_{plasma}}{\zeta_{untreated}}$ ratio different than unity may be indicative of pore blocking effects, or a modified pore size distribution.

**Conclusions**

We have shown that plasma processing can be used to modulate the capacitance as well as the resistance of AC electrodes prepared through powder consolidation. It was generally found that an optimal plasma processing is necessary for enhanced capacitance in the prepared electrochemical capacitors. Additional or excessive processing yields an overall reduction in the capacitance. A correlation of the parameters obtained standard electrochemical characterization techniques (CCCD, CV, and EIS) was found necessary to yield insights into the specific pore size medication and distribution effects. Considering the difficulty of representing actual porous matrices, and as an alternative to relatively simplified lumped parameter models, the product of the experimentally determined resistance and capacitance values along with the experimental testing time has been posited as a tool for analysis. The consequent parameterization of the electrolyte reach into the porous matrix in terms of the $\zeta$ may be a useful metric to correlate the extent to which a testing procedure may be considered complete.

**Acknowledgments**

We would like to thank the Nano3 Fabrication lab at UCSD. We appreciate the help of Dr. Seth Cohen and Dr. Zhenjie Zhang, in the Department of Chemistry at UC, San Diego in BET surface measurements and calculations. The authors acknowledge financial support from the Defense Advanced Research Projects Agency (DARPA: W911NF-15-2-0122) and the National Science Foundation (NSF: CMMI 1246800),

**References**

1. D. Linden and T. B. Reddy, *Linden's Handbook of Batteries*, McGraw Hill, New York, NY, (2010).

2. J. R. Miller and A. F. Burke, *Electrochem. Soc. Interface*, 53–57 (2008).

3. J. M. Miller, *Ultracapacitor Applications*, The Institution of Engineering and Technology, Herts, UK, (2011).

4. H. Shi, *Electrochim. Acta*, **41**, 1633 (1996).

5. P. Simon and Y. Gogotsi, *Acc. Chem. Res.*, **46**, 1094–1103 (2012)

6. D. Qu and H. Shi, *J. Power Sources*, **74**, 99–107 (1998).

7. B. E. Conway, *J. Electrochem. Soc.*, **138**, 1539–1548 (1991).

8. R. Narayanan and P. R. Bandaru, *J. Electrochem. Soc.*, **162**, A86–A91 (2015).

9. P. Simon and Y. Gogotsi, *Nat. Mater.*, **7**, 845–854 (2008).

10. A. Peigney, C. Laurent, E. Flahaut, R. R. Basca, and A. Rousset, *Carbon N. Y.*, **39**, 507–514 (2001).

11. P. R. Bandaru, H. Yamada, R. Narayanan, and M. Hoefer, *Mater. Sci. Eng. R Reports*, **96**, 1–69 (2015)

12. M. Hoefer and P. R. Bandaru, *J. Electrochem. Soc.*, **160**, H360–H367 (2013).

13. M. Hoefer and P. R. Bandaru, *J. Appl. Phys.*, **108**, 34308 (2010).

14. R. Narayanan et al., *Nano Lett.*, **15**, 3067–3072 (2015)

15. J. Nichols, C. P. Deck, H. Saito, and P. R. Bandaru, *J. Appl. Phys.*, **102**, 64306 (2007).

16. M. A. Hoefer and P. R. Bandaru, *J. Appl. Phys.*, **108**, 034308 (2010).

17. M. A. Hoefer, thesis, University of California, San Diego (2012).

18. K. Okajima, K. Ohta, and M. Sudoh, *Electrochim. Acta*, **50**, 2227–2231 (2005)

19. B. E. Conway, in *Electrochemical Supercapacitors- Scientific Fundamentals and Technological Applications,*, Kluwer- Academic (1999).

20. B. E. Conway, E. Gileadi, and M. Dzieciuch, *Electrochim. Acta*, **8**, 143–161 (1963).

21. G. Xu et al., *Nano Res.*, **4**, 870–881 (2011)

22. N. Sadeghi, M. van de Grift, D. Vender, G. M. W. Kroesen, and F. J. de Hoog, *Appl. Phys. Lett.*, **70**, 835–837 (1997).

23. M. D. Stoller and R. S. Ruoff, *Energy Environ. Sci.*, **3**, 1294 (2010)

24. M. E. Orazem and B. Tribollet, *Electrochemical Impedance Spectroscopy*, John Wiley Inc., New York, NY, (2008).

25. E. Barsoukov and J. R. MacDonald, *Impedance Spectroscopy: Theory, Experiment, and Applications*, Wiley-Interscience, Hoboken, NJ, (2005).

26. X. Andrieu, in *Energy Storage Systems for Electronics*, T. Osaka and M. Datta, Editors, p. 521–545, Gordon and Breach Science Publishers, Amsterdam (2000).

27. H. Keiser, K. D. Beccu, and M. A. Gutjahr, *Electrochim. Acta*, **21**, 539–543 (1976)

28. W. G. Pell and B. E. Conway, *J.Electroanal.Chem.*, **500**, 121–133 (2001).

29. J. Gamby, P. . Taberna, P. Simon, J. . Fauvarque, and M. Chesneau, *J. Power Sources*, **101**, 109–116 (2001).

30. R. de Levie, *Electrochim. Acta*, **9**, 1231–1245 (1964)

31. H.-K. Song, H.-Y. Hwang, K.-H. Lee, and L. H. Dao, *Electrochim. Acta*, **45**, 2241–2257 (2000)

32. R. de Levie, *Electrochim. Acta*, **8**, 751–780 (1963)

33. S. Ramo, J. R. Whinnery, and T. van Duzer, *Fields and Waves in Communication Electronics* J. W. & Sons, Editor, 3rd ed., New York, (1993).

**Figure Captions**

Figure 1

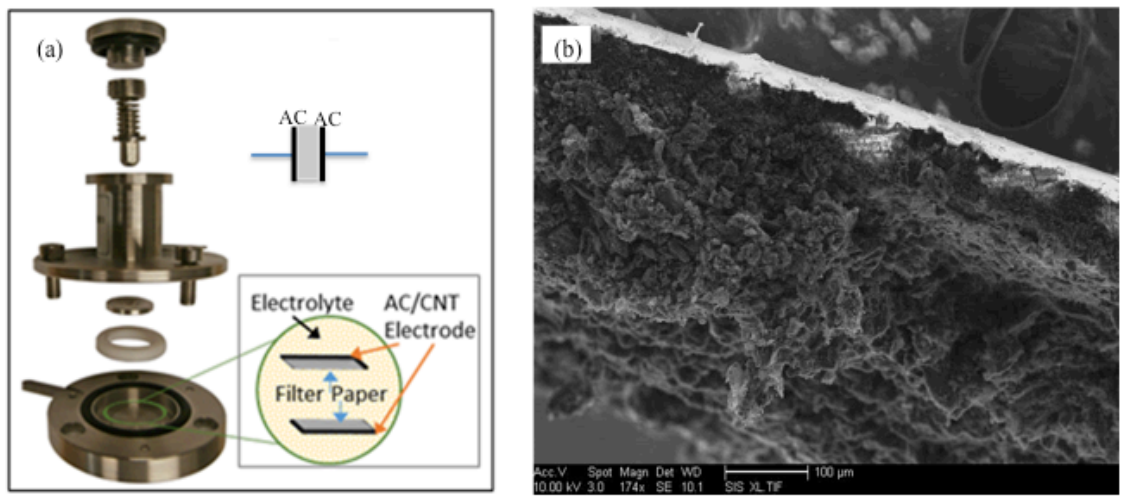

Marcelis, Narayanan, and Bandaru

**Figure 1.**

**(a)** Schematic of the setup for the activated carbon (AC) electrode based electrochemical capacitor (EC), constituted from two AC electrodes, and **(b)** Scanning electron microscope (SEM) image of the AC electrode.

**Figure 2**

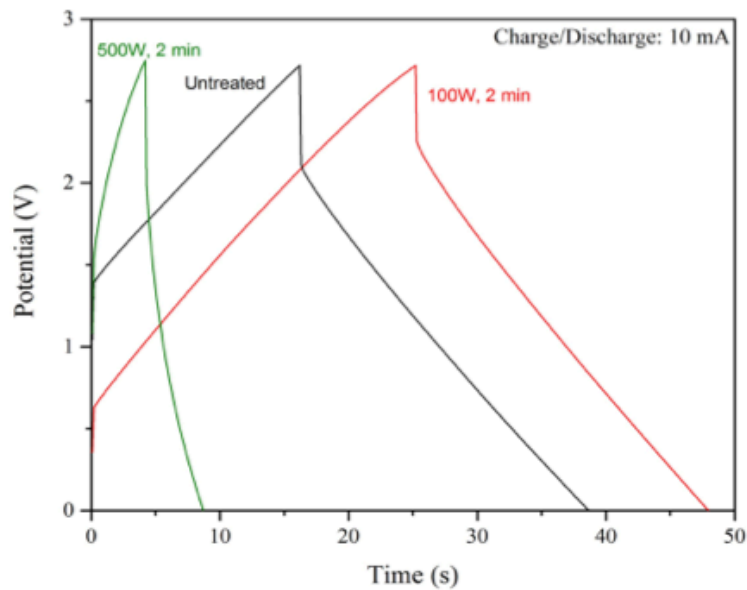

Marcelis, Narayanan, and Bandaru

**Figure 2.** Potential/voltage–time characteristics obtained during constant current charge/discharge (CCCD) based methodology used for the characterization of the AC electrodes as a function of plasma processing. A constant current ($I$) of 10 mA was applied in the charging (/discharging) of the EC. The equivalent resistance (= $R_{CCCD}$) and the capacitance (= $C_{CCCD}$) was obtained from the *IR* drop and the slope of the discharge characteristic.

**Figure 3**

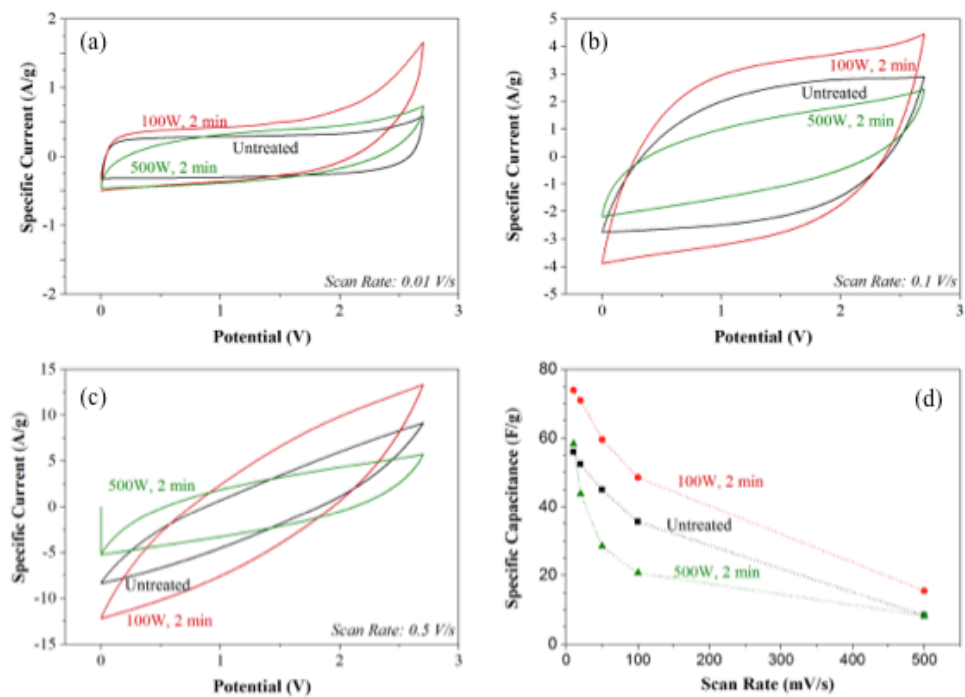

Marcelis, Narayanan, and Bandaru

**Figure 3.** Cyclic Voltammetry (CV) characteristics comparing untreated AC electrodes and plasma processed electrodes (100 W @ 2 minutes, and 500 W @ 2 minutes), at a scan rate (= $v$) of **(a)** 0.01 V/s, **(b)** 0.1 V/s, and **(a)** 0.5 V/s. **(d)** The general decrease of the specific capacitance with $v$ indicates diminishing access of the electrolyte ions into the pores. An overall general increase (/decrease) of the capacitance for the 100 W (/500 W) plasma processed samples is observed.

**Figure 4**

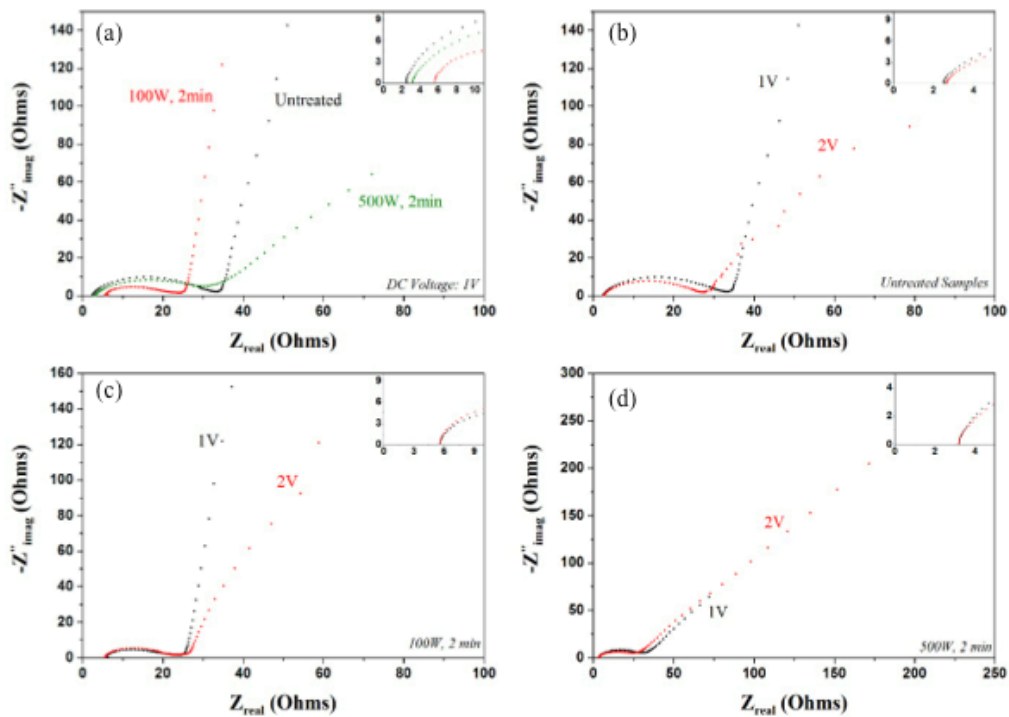

Marcelis, Narayanan, and Bandaru

**Figure 4.** Electrochemical Impedance Spectroscopy (EIS) was carried out in the range of 10 mHz to 300 kHz, on untreated and plasma processed electrodes. The presence and modulation of the semi-circular loop at high frequencies, indicates variable pore morphology. The EIS spectra, **(a)** comparing untreated, with the plasma processed (@ 100 W and 500 W) samples; at *inset:* at the highest frequencies, considering the intercept on the real axis, it is indicated that the series resistance is enhanced due to the plasma processing. EIS spectra as a function of DC bias for the **(b)** untreated AC electrodes, **(c)** 100 W plasma processed electrodes, and the **(d)** 500 W plasma processed electrodes. The insets in (b), (c), and (d) indicate the spectra at the highest frequencies in each case. The dispersion of the capacitance at low frequencies indicates diffusional behavior, and may be related to pore size distribution.

Figure 5

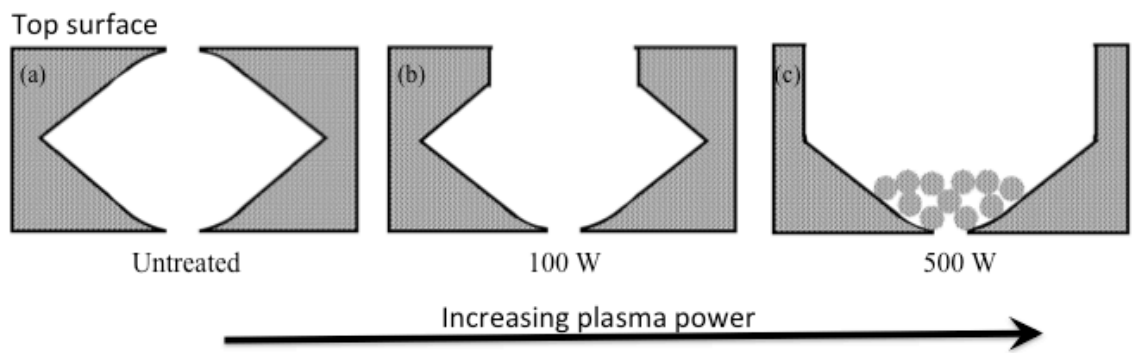

Marcelis, Narayanan, and Bandaru

**Figure 5.** The hypothesized variation of a *typical* pore for **(a)** untreated, **(b)** the 100 W plasma processed, and the **(c)** 500 W plasma processed AC electrodes. The basis pore size in (a) was deduced from the correlation of the EIS spectra[27].

**Table I**

The gravimetric capacitances of the untreated, and plasma processed AC electrodes (for 100 W @ 2 minutes, and 500 W @ 2 minutes), with respect to three electrochemical characterization procedures, *i.e.,* CCCD (Constant current charge/discharge), CV (cyclic voltammetry), and EIS (electrochemical impedance spectroscopy)

|      | Untreated | 100 W @ 2 min | 500 W @ 2 min |
|------|-----------|---------------|---------------|
| CCCD | 25.6 F/g  | 31.6 F/g      | 8.8 F/g       |
| CV   | 35.6 F/g  | 48.4 F/g      | 20.8 F/g      |
| EIS  | 28.5 F/g  | 34.3 F/g      | 15.2 F/g      |

**Table II**

The variation of the experimentally determined resistances ($R_{det}$) and capacitance ($C_{det}$); $det$ = *CCCD*, *CV*, or *EIS*, for untreated and plasma processed samples. The *reach* ($\zeta = \sqrt{t \cdot \tau}$) of the electrolyte, for each determination and sample, is characterized by the product of the experimental time ($t$) as well as a time constant, $\tau = R_{det}C_{det}$.

|  | CCCD | | | CV | | | EIS | | |
| --- | --- | --- | --- | --- | --- | --- | --- | --- | --- |
|  | Untreated | 100 W | 500 W | Untreated | 100 W | 500 W | Untreated | 100 W | 500 W |
| $t$ | 22.0 s | 22.0 s | 5.0 s | 27.0 s | 27.0 s | 27.0 s | 105.0 s | 105.0 s | 105.0 s |
| $R_{det}$ | 30 Ω | 23 Ω | 38 Ω | 46 Ω | 43 Ω | 130 Ω | 34 Ω | 26 Ω | 24 Ω |
| $C_{det}$ | 105 mF | 101 mF | 22 mF | 146 mF | 155 mF | 52 mF | 117 mF | 110 mF | 38 mF |
| $\tau$ | 3.2 s | 2.3 s | 0.8 s | 6.7 s | 6.7 s | 6.7 s | 4.0 s | 2.9 s | 0.9 s |
| $\zeta$ | 8.4 s | 7.3 s | 2.0 s | 13.5 s | 13.5 s | 13.5 s | 20.5 s | 17.2 s | 9.7 s |